# Experimental Demonstration of Chiral Bound States in the Continuum Enabled by Time-reversal Symmetry Breaking

Hao-Chang Mo[#], Wen-Jin Zhang[#], Ze-Yu Wu, Xin-Tao He, Xiao-Dong Chen[*], Jian-Wen Dong

*School of Physics & State Key Laboratory of Optoelectronic Materials and Technologies, Sun Yat-sen University, Guangzhou 510275, China.*

[#]*The authors contributed equally to this work*

[*]Corresponding author: chenxd67@mail.sysu.edu.cn

**Bound states in the continuum (BICs) provide a powerful route to optical resonances with vanishing radiation loss and enhanced light-matter interactions. Of particular interest are chiral BICs, whose resonant states exhibit intrinsic circular polarization and spin-selective radiation. Magneto-optical photonic crystals have recently been predicted to host such states through time-reversal-symmetry breaking. Here, we experimentally demonstrate chiral BICs induced by time-reversal symmetry breaking in a magneto-optical photonic crystal, where an out-of-plane magnetic field lifts a degenerate BIC pair into two nondegenerate resonances with opposite circular polarizations. Numerical simulations reveal the associated chiral phase vortices and strong spin-selective radiation, while microwave measurements confirm the predicted behavior through angle-resolved transmission spectra and pronounced broadband circular dichroism. Our results establish an experimental platform for exploring chiral BIC physics in magneto-optical photonic structures, opening new opportunities for spin-selective photonics.**

***Introduction.*** --- Bound states in the continuum (BICs) represent a class of wave phenomena in which spatially localized modes coexist with a continuum of radiative states while remaining perfectly confined [1-5]. First proposed in quantum mechanics, the concept of BICs has attracted extensive attention in photonics because it provides a powerful mechanism for realizing resonances with theoretically diverging quality factors [6-14]. In photonic systems, BICs arise when radiation channels are suppressed by destructive interference [2,5,13] or symmetry mismatch with free-space modes [3,14,15], enabling electromagnetic energy to remain localized despite the presence of available radiation pathways. Periodic photonic crystals provide particularly versatile platforms for realizing BICs, where such states typically appear at specific points in momentum space where coupling between resonant modes and radiation channels vanishes [3,10,13,15-17]. A representative example is the symmetry-protected BIC at the center of the Brillouin zone (Γ point), where certain modes cannot radiate because of symmetry incompatibility with free-space modes. When the protecting symmetry is slightly perturbed, ideal BICs evolve into quasi-BICs with extremely high but finite quality factors [13,18-20], which have been widely exploited to realize narrow resonances and enhanced light-matter interactions in photonic devices [21-25].

Particular interest has recently been devoted to chiral BICs and chiral quasi-BICs, where resonant modes exhibit circular polarization and spin-selective radiation [26-38]. Existing routes to such states have mainly relied on structural symmetry engineering, for example by breaking the out-of-plane mirror symmetry [27,33], introducing in-plane inversion asymmetry [28,34], combining in-plane and out-of-plane asymmetry [29,35], or exploiting interlayer coupling in twisted or bilayer photonic crystal slabs [36,37], all of which can convert symmetry-protected BICs into leaky chiral quasi-BICs or arbitrarily polarized modes associated with BICs. Beyond structural engineering, active tuning of BICs and their associated polarization singularities has also been explored using optically excited semiconductors [39], electrically tunable two-dimensional materials [40], phase-change materials [41], and tunable plasma platforms [42]. These approaches provide additional material degrees of freedom for dynamically controlling BIC resonances and the evolution of polarization singularities without necessarily modifying the structural geometry. In parallel, magneto-optical photonic crystals (MOPCs) provide a particularly distinct route [31,38], where an external magnetic field breaks time-reversal symmetry, thereby lifting a degenerate BIC pair into circularly polarized eigenstates with opposite chirality. This mechanism has recently been predicted to generate chiral BICs, robust intrinsic C points, and magneto-optical BICs in several theoretical works. However, although the underlying physics of time-reversal-symmetry-breaking chiral BICs has been theoretically established, their direct experimental observation in magneto-optical photonic structures remains largely unexplored.

Here, we experimentally demonstrate chiral BICs induced by time-reversal symmetry breaking in an MOPC operating at microwave frequencies. In the absence of an external magnetic field, the system supports a degenerate BIC pair at the Γ point. When an out-of-plane magnetic field is applied, time-reversal symmetry is broken and the degeneracy is lifted into two nondegenerate chiral BICs with opposite circular polarizations. Numerical simulations reveal the associated chiral phase vortices and spin-selective radiation, and microwave measurements further confirm these features through angle-resolved transmission spectra and pronounced circular dichroism (CD). These results establish an experimental platform for exploring chiral BIC physics in magneto-optical photonic structures and suggest new opportunities for spin-selective photonics.

***Chiral BICs under time-reversal symmetry breaking.*** --- We first analyze the physical mechanism responsible for the formation of degenerate (chiral) BICs in the MOPC under preserved (broken) time-reversal symmetry. As illustrated schematically in Fig. 1(a), the system consists of periodically arranged meta-atoms forming a two-dimensional honeycomb lattice, whose Bloch modes are defined within the first Brillouin zone shown in Fig. 1(b). To enhance the magneto-optical interaction while maintaining a nearly uniform out-of-plane magnetic bias, a double "Big Mac" structure is adopted, in which the two yttrium iron garnet (YIG) rods are sandwiched between metallic disks or permanent magnets, as shown in Fig. 1(c). In the absence of an external magnetic field, the system preserves time-reversal symmetry. Under this condition, two parabolic photonic bands become degenerate at the Γ point, as represented by the leftmost cross-sectional plane in Fig. 1(d). At the band intersection (black dot), two degenerate states form symmetry-protected degenerate BICs due to their incompatibility with free-space radiation channels.

When an external magnetic field is applied along the $z$ direction [right panel of Fig. 1(c)], the gyromagnetic response of the YIG rods is induced and the off-diagonal components of the permeability tensor become nonzero. This magneto-optical coupling breaks the time-reversal symmetry and lifts the degeneracy of the BIC pair. As a result, the two linearly polarized eigenmodes evolve into two nondegenerate circularly polarized states with opposite handedness. The corresponding evolution with increasing time-reversal-symmetry-breaking strength $\delta$ is schematically illustrated in Fig. 1(d): the double degeneracy is lifted, splitting the initial state into a pair of intrinsically chiral BICs, i.e., modes that fundamentally correspond to opposite circular polarizations but exhibit zero far-field radiation. Here, the highly symmetric structure is deliberately retained so that the nonradiative symmetry-protected BIC pair at the Γ point serves as a well-defined parent state. This allows the evolution from two degenerate orthogonal linearly polarized modes to two non-degenerate circularly polarized modes with opposite handedness to be attributed unambiguously to magneto-optical coupling and time-reversal-symmetry breaking, rather than to structural symmetry breaking. Introducing a controlled structural perturbation, such as Brillouin-zone folding, would convert these ideal chiral BICs into radiative chiral quasi-BICs with finite $Q$ factors, providing a natural route toward directly excitable device implementations.

***Numerical demonstration of chiral BICs.*** --- To verify the above physical picture, we perform full-wave numerical simulations of the MOPC. Fig. 2(a) demonstrates the bulk bands of the structure composed of meta-atoms (two YIG cylinders with a diameter of $d$ = 5.6 mm and a height of $h_1$ = 3 mm, and three metallic disks with a diameter $d$ = 5.6 mm and a height of $h_2$ = 1 mm) that are arranged in a honeycomb lattice with a lattice constant of $a$ = 14 mm. When the time-reversal symmetry is preserved, two degenerate modes appear at the Brillouin zone center at 10.83 GHz. These two modes constitute a degenerate BIC pair, since their symmetries are incompatible with those of the free-space radiation channels, as illustrated by the field profiles in the right panels of Fig. 2(b). For these two degenerate BIC modes, the phase of the out-of-plane electric-field component $E_z$ is predominantly restricted to two values differing by π, resulting in the parallel/antiparallel arrow patterns. The BIC nature of these modes is further confirmed by their k-space Q-factor distributions (see Sec. S4 of the Supplemental Material). Away from the Γ point, the bulk modes gradually become radiative, and their far-field polarization distributions are shown in the left panels of Fig. 2(b). Notably, these radiative states on the two bands remain linearly polarized and mutually orthogonal, indicating that the photonic crystal does not exhibit intrinsic chirality when the time-reversal symmetry is preserved.

The situation changes dramatically once an out-of-plane magnetic field is applied and magneto-optical coupling is introduced. As shown in Fig. 2(c), the degenerate BIC pair at the Γ point splits into two nondegenerate resonances. The $E_z$ field profiles shown in the right panels of Fig. 2(d) exhibit opposite spatial phase windings around the YIG cylinders, manifested as vortex and antivortex phase textures, respectively. These opposite phase windings provide a real-space signature distinguishing the two chiral states. Meanwhile, the far-field polarization distributions shown in the left panels of Fig. 2(d) reveal that the radiative states around the two split bands exhibit predominantly left-circular polarization (LCP, σ−) and right-circular polarization (RCP, σ+), respectively, confirming the opposite chiralities of the two chiral BICs in sharp contrast to the linearly polarized states in the time-reversal-symmetric case. The spin-selective response associated with these chiral BICs is further examined through the simulated transmission spectra at a small incidence angle of 10°, as shown in Fig. 2(e). Here, the small-angle incidence serves as a probe of the nonradiative chiral BICs: the modes retain nearly the same chiral polarization as those at the Γ point while becoming weakly radiative, allowing their circular-polarization selectivity to be observed in the transmission spectra. The pronounced difference between $T_{\mathrm{RR}}$ and $T_{\mathrm{LL}}$ shows that the two chiral BIC bands selectively couple to opposite circular polarizations. Specifically, the lower- and upper-frequency resonances originate from the lower and upper bands associated with the chiral BIC pair (approximately 9.75 and 10.8 GHz, respectively), whereas the intermediate resonance arises from an additional spectator band (approximately 10.5 GHz). This spectator band is not involved in the formation or magnetically induced splitting of the chiral BIC pair and is therefore not displayed in Fig. 2(c) [see Sec. S5 and Fig. S4 of the Supplemental Material]. This feature is captured more directly in the circular dichroism spectrum in Fig. 2(f), where a pronounced and broadband CD response emerges. In the lower-frequency range from about 9.5 to 10.2 GHz, the CD remains close to +1, indicating an almost complete preference for one handedness. In contrast, the CD reverses sign and approaches −1 in the higher-frequency range from approximately 10.5 to 11.0 GHz, corresponding to an opposite handedness selectivity. This broadband evolution from positive to negative near-unity CD provides clear numerical evidence for the formation of chiral BICs induced by time-reversal symmetry breaking and highlights their strong chiral radiation response over a wide frequency range.

***Experimental observation of chiral BICs.*** --- To experimentally validate the predicted chiral BICs, we fabricated an MOPC sample composed of YIG cylinders together with either metallic disks or permanent magnets, corresponding to the time-reversal-symmetric and symmetry-broken configurations, respectively, as shown in Fig. 3(a). A close-up view of the fabricated sample is shown in Fig. 3(b), where the periodic honeycomb arrangement of the meta-atoms can be clearly identified. Each meta-atom, which consists of two YIG cylinders sandwiched between three metallic disks or permanent magnets, is embedded in a low-loss dielectric foam substrate (ROHACELL 31HF) for mechanical support. When metallic disks are used, the YIG cylinders remain unbiased and the system preserves time-reversal symmetry. By contrast, when permanent magnets are used, they provide a static out-of-plane magnetic field that magnetizes the YIG cylinders and induces the magneto-optical coupling required to break time-reversal symmetry. The experimental geometric parameters of the fabricated sample, including the lattice constant and the height and diameter of the YIG cylinders, are chosen to match those used in the numerical simulations. In particular, magnets with a diameter of 5.6 mm and a height of 1 mm are employed to provide the required out-of-plane magnetic bias field $B_0$ = 0.058 T. As shown in Fig. 3(c), all experiments are performed using a customized angle-resolved far-field measurement platform (Linbou FFS). The sample is mounted on a rotator that automatically varies the polar angle $\theta$. A pair of highly collimated lens antennas (HD-100LHA250, operating from 8.2 to 12.4 GHz) is used for transmission and reception and mounted on two long arms separated by approximately 1500 mm. The polarization of each antenna is adjusted by rotating the entire antenna about its propagation axis. For the *s*-polarized (*p*-polarized) configuration, the short sides are perpendicular (parallel) to the ground, as shown in Fig. 3(d, e). All antennas are connected to a vector network analyzer (VNA; Keysight E5071C, operating from 0.3 MHz to 20 GHz) to measure transmission spectra. During the experiment, the rotator is controlled by computer software to rotate automatically from −50° to +50°. When the rotator stops at a predetermined angle $\theta$, a linearly polarized wave is launched from the transmitting antenna toward the sample, and the transmitted wave is collected by the receiving antenna. To characterize the circular-polarization response, the four complex transmission coefficients in the linear-polarization basis, $t_{ss}$, $t_{sp}$, $t_{ps}$ and $t_{pp}$, are measured by changing the orientations of the transmitting and receiving antennas. Here, $t_{\alpha\beta}$ denotes the complex transmission coefficient detected in the $\alpha$-polarized channel under $\beta$-polarized incidence. Since the vector network analyzer records both the amplitude and phase of these complex coefficients, the LCP and RCP transmission components can be reconstructed through a linear-to-circular polarization basis transformation in post-processing, in which the required $\pm\pi/2$ relative phase between the two orthogonal linear components is introduced mathematically rather than by an external phase shifter. The co-polarized LCP and RCP transmission coefficients are given by

$$t_{\mathrm{LL}} = \frac{1}{2}\left[\left(t_{pp} + t_{ss}\right) - i\left(t_{ps} - t_{sp}\right)\right] \text{ and } t_{\mathrm{RR}} = \frac{1}{2}\left[\left(t_{pp} + t_{ss}\right) + i\left(t_{ps} - t_{sp}\right)\right] \tag{1}$$

The circular dichroism is defined as

$$\mathrm{CD} = \frac{T_{\mathrm{RR}} - T_{\mathrm{LL}}}{T_{\mathrm{RR}} + T_{\mathrm{LL}}} \tag{2}$$

where $T_{\mathrm{LL}} = |t_{\mathrm{LL}}|^2$ and $T_{\mathrm{RR}} = |t_{\mathrm{RR}}|^2$ denote the co-polarized LCP and RCP transmittances, respectively. Accordingly, negative (positive) CD corresponds to stronger LCP (RCP) transmission.

We now present the experimental results demonstrating the existence of chiral BICs. Because the ideal BICs at the Γ point are decoupled from free-space radiation, they are not identified experimentally by directly measuring an infinitely narrow resonance at normal incidence. Instead, their existence and chirality are inferred from the magnetically induced band splitting, the opposite circular-polarization selectivity of the nearby radiative bands, and their continuous evolution toward the nonradiative states at the Γ point. We first consider the case without time-reversal symmetry breaking, where the metallic disks are used and the YIG cylinders remain unbiased. As shown in Fig. 4(a), the measured angle-resolved transmission spectra exhibit two bands near the Γ point, in good agreement with the numerical simulations. To further identify the polarization characteristics of the two modes, we analyze the transmission spectra under different linear polarization channels. The upper band is primarily revealed in the $T_{pp}$ (= $|t_{pp}|^2$) spectrum, whereas the lower band appears dominantly in the $T_{ss}$ (= $|t_{ss}|^2$) spectrum, indicating their respective coupling to $x$- and $y$-polarized excitations. This observation is fully consistent with the numerical results in Fig. 2, confirming that, under preserved time-reversal symmetry, the two bands are selectively excited through orthogonal linear polarization waves. The situation changes markedly when the metallic disks are replaced by permanent magnets, so that an out-of-plane magnetic bias is introduced and time-reversal symmetry is broken. In this case, the degenerate bands are clearly split [Fig. 4(b)]. Simultaneously, the magneto-optical interaction hybridizes the originally orthogonal linear states, giving rise to two chiral bands with opposite handedness. Specifically, the upper band corresponds predominantly to a left-circularly polarized state, while the lower band corresponds predominantly to a right-circularly polarized state. This chiral selectivity is directly confirmed by polarization-resolved transmission measurements: for the response reconstructed in the LCP basis, only the upper band is efficiently excited [left panel of Fig. 4(b)], whereas in the response reconstructed in the RCP basis, the lower band dominates [right panel of Fig. 4(b)].

The corresponding transmission spectra at $\theta = 35°$ along the Γ-K path further reveal a strong spin-selective response. As shown in Fig. 4(c), in the higher-frequency range from approximately 10.2 to 10.8 GHz, the transmitted left-circularly polarized component is much stronger than the right-circularly polarized one. In contrast, in the lower-frequency window from approximately 9.0 to 10.2 GHz, the situation is reversed, with the right-circularly polarized transmission becoming dominant. This reversal is more clearly captured by the corresponding circular dichroism spectrum in Fig. 4(d), where the CD evolves from nearly +1 to nearly −1 across the two frequency regions. Importantly, the measured CD remains close to its extreme values over broad frequency windows, demonstrating a pronounced broadband chiral response. The metallic disks or permanent magnets are essential for the formation of this high-CD window. They generate a strongly frequency-dependent reflection background with pronounced transmission edges, which are shifted in opposite directions for RCP and LCP by the gyromagnetic coupling (see Sec. S3 of the Supplemental Material for details). These experimental results agree well with the numerical simulations and provide clear evidence for the formation of chiral BICs induced by time-reversal symmetry breaking.

***Conclusion and Discussion.*** --- We have experimentally demonstrated chiral BICs induced by time-reversal symmetry breaking in an MOPC. In the absence of an external magnetic field, the structure supports a degenerate BIC pair at the Γ point composed of two orthogonal linearly polarized modes. When an out-of-plane magnetic field is applied, magneto-optical coupling lifts this degeneracy and transforms the original BIC pair into two nondegenerate resonances carrying opposite circular polarizations. Numerical simulations further reveal the associated chiral phase vortices and spin-selective radiation, while microwave experiments confirm these features through angle-resolved transmission spectra and pronounced broadband circular dichroism.

Recent experiments have revealed several complementary manifestations of chiral BICs in magneto-optical photonic crystals, including self-biased spin–orbit-locked chiral BICs arising from magnetically lifted BIC degeneracies [43], single-band intrinsic planar chiral BICs generated through the interplay of spatial-symmetry and time-reversal-symmetry breaking [44], and multidimensional real-space topological structures associated with chiral BICs [45]. Within this emerging experimental landscape, our results not only provide a direct and field-tunable realization of the transition from a degenerate BIC pair to oppositely handed chiral resonances but also establish a direct experimental platform for investigating chiral BIC physics in magneto-optical photonic structures. In contrast to conventional structural routes for generating chiral resonances, the present approach directly exploits time-reversal symmetry breaking as the key degree of freedom, providing a physically transparent route from BIC degeneracy to chirality. Such a mechanism not only preserves the high-$Q$ nature inherited from BIC physics, but also enables strong and broadband circular-polarization selectivity in the radiative response. More broadly, the combination of magneto-optical coupling and BIC resonances may open new possibilities for spin-selective photonics, magnetically controlled chiral responses, and nonreciprocal photonic devices based on high-$Q$ resonant states.

**Acknowledgements**

This work was supported by National Natural Science Foundation of China (12522415, 12374364, 12274475), Guangdong Basic and Applied Basic Research Foundation (2023B1515040023, 2023B1515020072), Guangzhou Science, Technology and Innovation Commission (2024A04J6333). In addition, we gratefully acknowledge the Modern Matter Laboratory and Advanced Materials Thrust at The Hong Kong University of Science and Technology (Guangzhou) for providing experimental support. We particularly thank Prof. Xiaoxiao Wu and Dr. Haitao Li for their assistance with the measurements.

## Figures and Captions

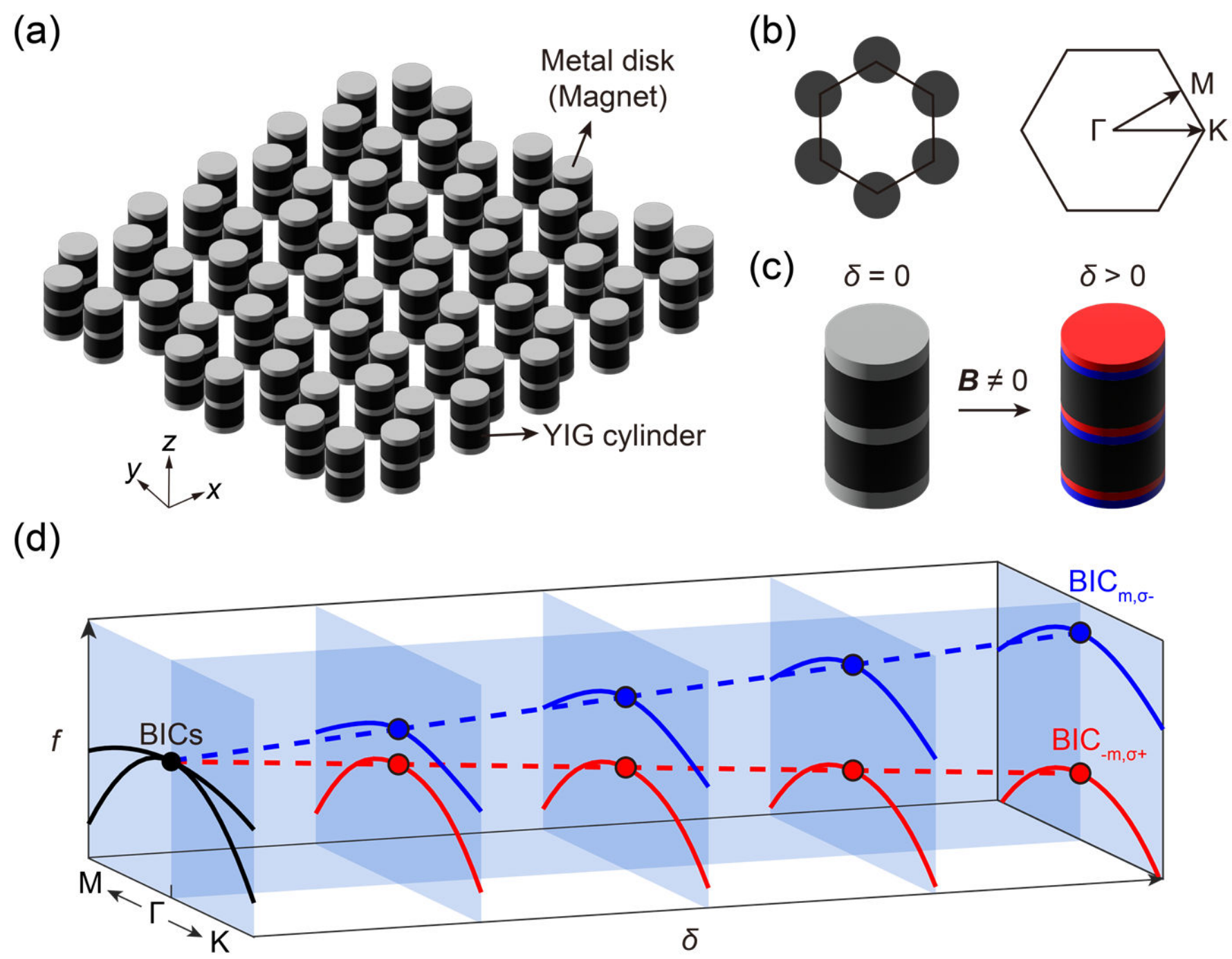

**FIG. 1. Chiral BICs under time-reversal symmetry breaking.** (a) Schematic illustration of the MOPC, consisting of periodically arranged meta-atoms in a honeycomb lattice. (b) Geometry of the photonic crystal unit cell (left) and the corresponding first Brillouin zone (right). (c) Meta-atom configuration without (left) and with (right) an out-of-plane magnetic field $B_0$. The external magnetic field induces magneto-optical coupling and breaks time-reversal symmetry. To simultaneously enhance the magneto-optical interaction and maintain a nearly uniform magnetic bias across the active region, a double "Big Mac" architecture is adopted, in which two YIG rods are sandwiched between metal disks or permanent magnets. (d) Schematic evolution of the resonant states at the Γ point as a function of the time-reversal-symmetry-breaking strength $\delta$. Without time-reversal symmetry breaking, two orthogonal modes form a degenerate BIC pair. As $\delta$ increases, the degeneracy is lifted and the BIC pair splits into two chiral BICs with opposite circular polarizations.

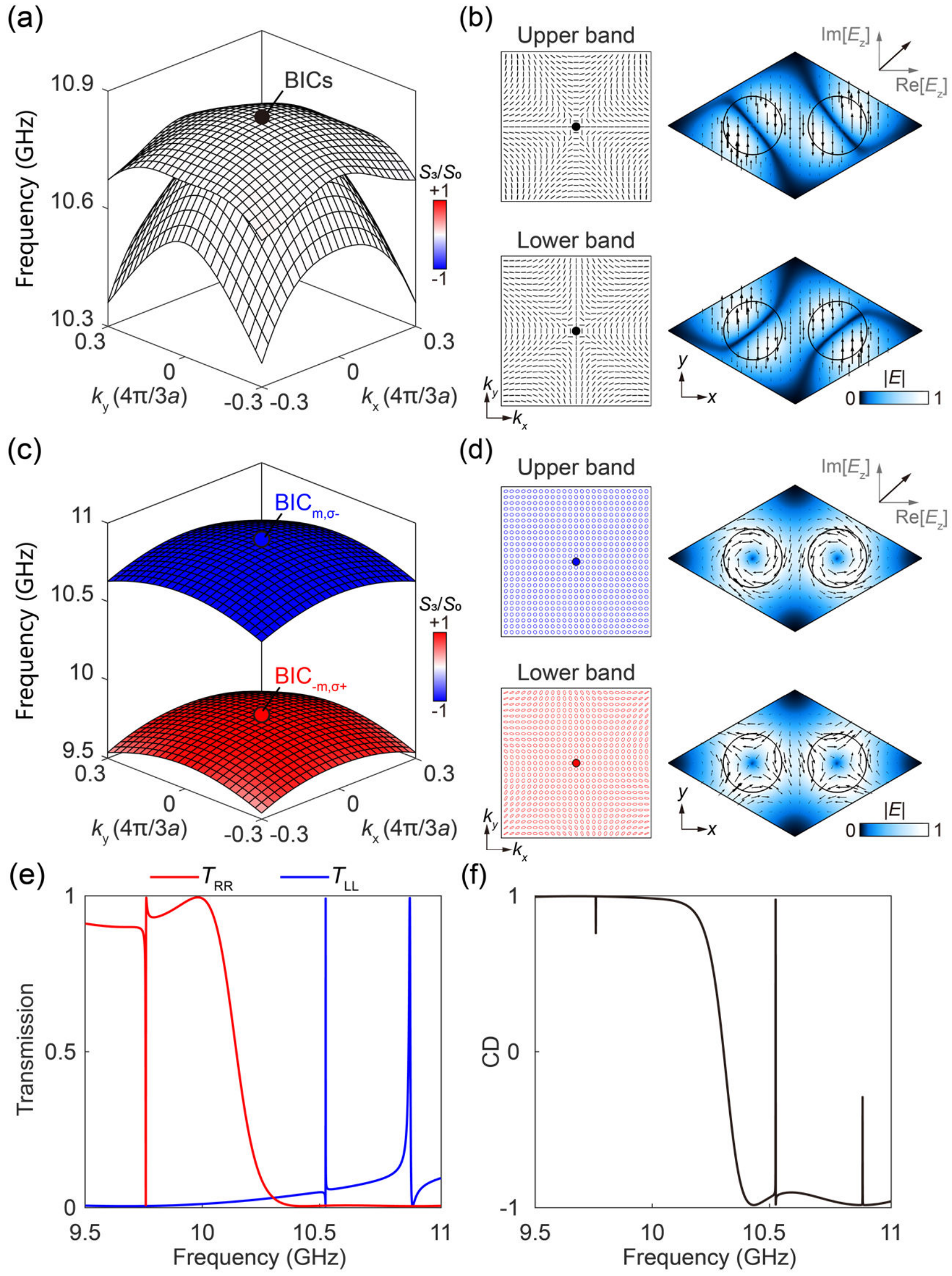

**FIG. 2. Numerical demonstration of chiral BICs.** (a) Bulk bands near the Γ point, showing a degenerate BIC pair at 10.83 GHz. (b) Far-field polarization distributions of the bulk modes around the Γ point (left), together with the electric-field profiles of the degenerate modes at the Γ point (right). The background color denotes the normalized electric-field magnitude $|E|$. The arrows represent the local complex phase of the out-of-plane electric-field component $E_z$, with their horizontal and vertical components corresponding to $\mathrm{Re}(E_z)$ and $\mathrm{Im}(E_z)$, respectively. (c) Bulk bands showing the splitting of the degenerate BIC pair into two nondegenerate resonances under magneto-optical coupling. (d) Far-field polarization distributions of the split bands (left) and the corresponding electric-field profiles at the Γ point (right), revealing two circularly polarized states with opposite chiral phase vortices. (e) Simulated co-polarized circular transmittances $T_{RR}$ and $T_{LL}$ at an incidence angle of 10°. The cross-polarized components $T_{RL}$ and $T_{LR}$ are zero throughout the frequency range considered. (f) Corresponding circular dichroism (CD) spectrum, exhibiting a pronounced broadband chiral response: the CD remains close to +1 in the lower-frequency range and reverses to nearly −1 in the higher-frequency range, confirming the emergence of chiral BICs induced by time-reversal symmetry breaking.

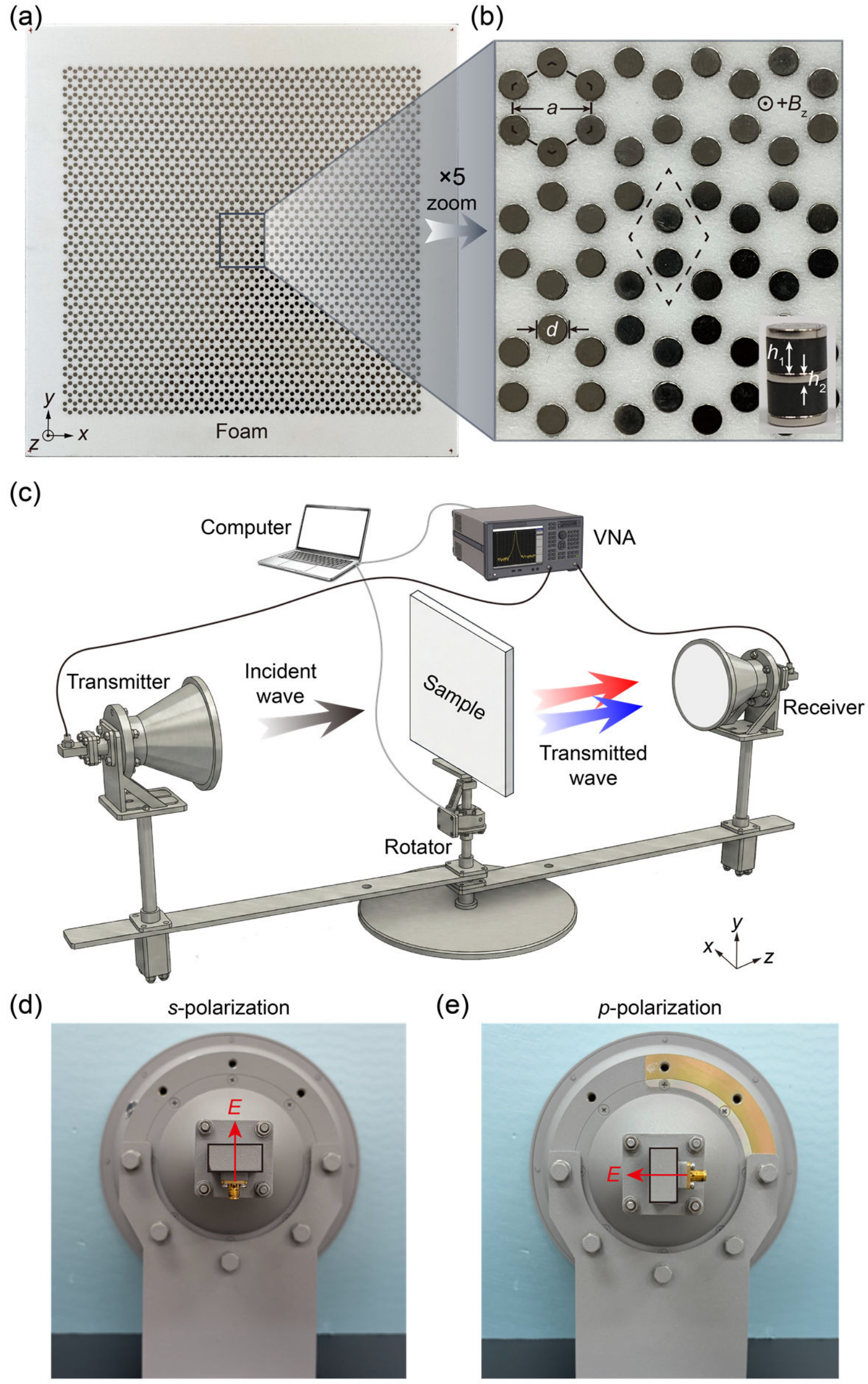

**FIG. 3. Experimental sample and measurement setup.** (a) Photograph of the fabricated photonic-crystal sample mounted on a low-loss foam substrate. (b) Enlarged view of the sample showing the periodic honeycomb arrangement of the meta-atoms. The inset shows a single meta-atom composed of two YIG cylinders sandwiched between three metallic or magnetic disks. When metallic disks are used, the YIG cylinders remain unbiased and the system preserves time-reversal symmetry; when permanent magnets are used, they provide the out-of-plane magnetic bias required for magneto-optical coupling and time-reversal-symmetry breaking. (c) Schematic of the microwave measurement setup. The sample is placed between transmitting and receiving horn antennas connected to a vector network analyzer for polarization-resolved transmission measurements. (d, e) The settings for *s*-polarized (d) and *p*-polarized (e) lens antennas. For the *s*-polarized (*p*-polarized) configuration, the short sides are perpendicular (parallel) to the ground.

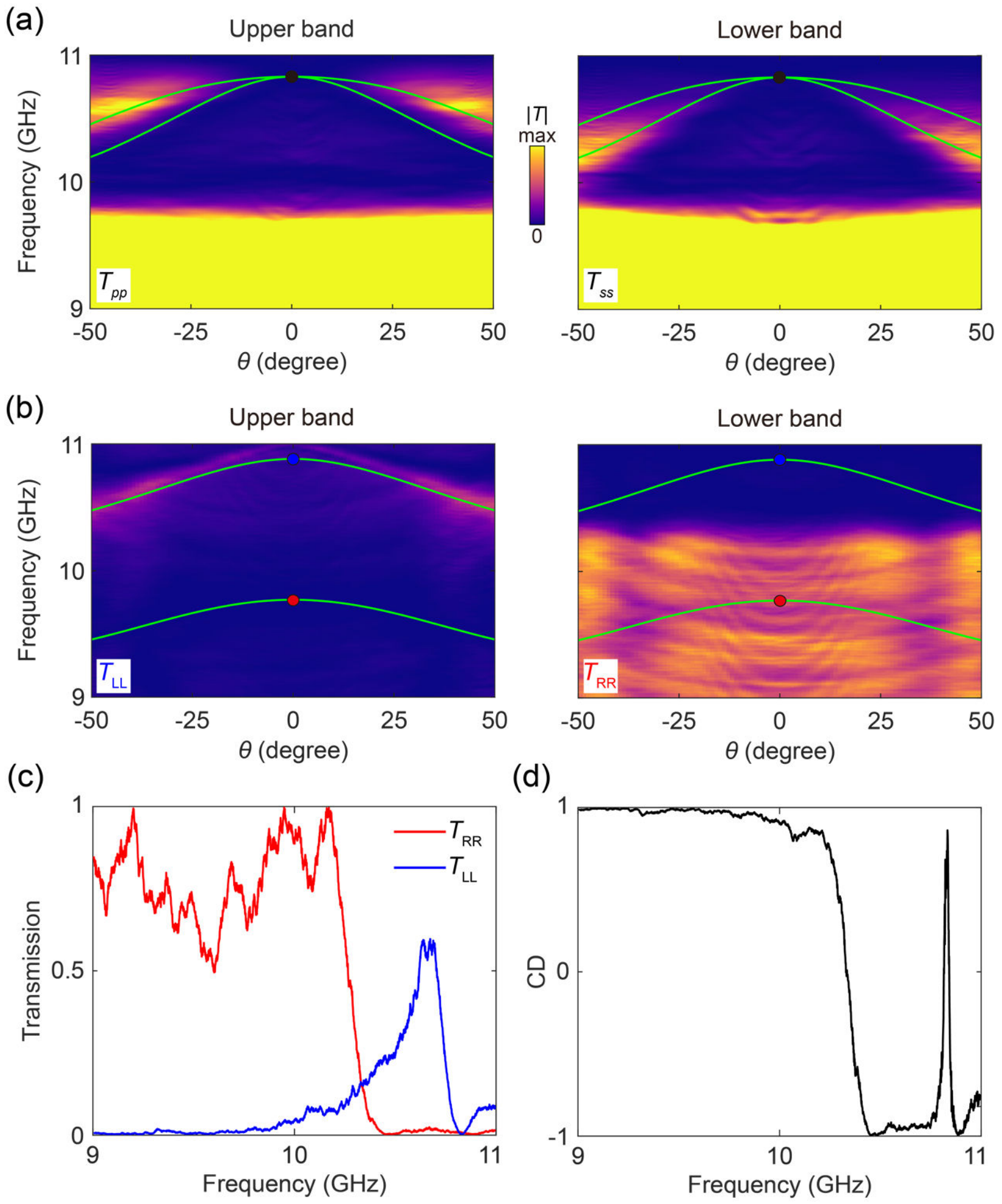

**FIG. 4. Experimental observation of chiral BICs.** (a) Measured angle-resolved transmission spectra without time-reversal-symmetry breaking. The green curves indicate the simulated band structure. By linear-polarization analysis of the transmitted waves, the upper and lower bands are identified as predominantly $x$- and $y$-polarized, respectively, consistent with the degenerate BIC pair composed of two orthogonal linearly polarized modes. (b) Measured angle-resolved transmission spectra transformed to the circular-polarization basis, shown for equivalent LCP-basis (left panel) and RCP-basis (right panel) excitations. The degenerate BIC pair is split into two nondegenerate bands, where the upper band is selectively excited by LCP waves and the lower band by RCP waves, indicating the emergence of opposite chirality. (c) Measured transmission spectra of the right- and left-circularly polarized output components at $\theta = 35°$ along the Γ-K path. In the lower-frequency range, the RCP transmission dominates over the LCP one, whereas the opposite behavior is observed in the higher-frequency range, revealing strong spin-selective radiation. (d) Corresponding circular dichroism spectrum, showing a pronounced broadband chiral response. The CD remains close to +1 in the lower-frequency band and reverses to nearly −1 in the higher-frequency band, in good agreement with the numerical simulations.